\documentclass[a4paper,12pt,reqno,superscriptaddress]{revtex4}
\usepackage[centertags]{amsmath}
\usepackage{amsfonts}
\usepackage{amssymb}
\usepackage{amsthm}
\usepackage{newlfont}
\usepackage{stmaryrd}
\usepackage{mathrsfs}
\usepackage{mathtools}
\usepackage{euscript}
\usepackage{graphicx}
\usepackage{enumerate}
\usepackage{todonotes}
\usepackage[normalem]{ulem} 

\usepackage{color}
\usepackage{floatrow}
\usepackage{caption}

\usepackage{tikz}
\usepackage{pgf}
\usetikzlibrary{positioning,fit,calc}
\usetikzlibrary{arrows,automata}
\usepackage{wrapfig}
\usepackage{subfigure}
\usepackage{amscd}
\usepackage{hyperref}

\theoremstyle{plain}

\theoremstyle{definition}

\theoremstyle{remark}

\newcommand{\opunit}{\text{1}\kern-0.22em\text{l}}

\DeclareMathAlphabet{\mathpzc}{OT1}{pzc}{m}{it}

\newcommand{\id}{\textrm{d}}

\let\oldsqrt\sqrt
\def\sqrt{\mathpalette\DHLhksqrt}
\def\DHLhksqrt#1#2{%
\setbox0=\hbox{$#1\oldsqrt{#2\,}$}\dimen0=\ht0
\advance\dimen0-0.2\ht0
\setbox2=\hbox{\vrule height\ht0 depth -\dimen0}%
{\box0\lower0.4pt\box2}}

\begin{document}

\title{Fluctuating motion in an active environment}

\author{Christian Maes\\
{\it Instituut voor Theoretische Fysica, KU Leuven}}

\begin{abstract}
We derive the fluctuation dynamics of a probe in weak coupling with a {\it living} medium, modeled as particles undergoing an active Ornstein-Uhlenbeck dynamics.       Nondissipative corrections to the fluctuation-dissipation relation are written out explicitly in terms of time-correlations in the active medium.
A first term changes the inertial mass of the probe, as a consequence of the persistence of the active medium.  A second correction modifies the friction kernel.  The resulting generalized Langevin equation benchmarks the motion induced on probes immersed in active {\it versus} passive media.  The derivation uses nonequilibrium response theory.
\end{abstract}
\maketitle


\section{Introduction}
For more than a century the paradigm of Brownian
motion and its variations have been used in mesoscopic physics to
describe fluctuating dynamics. Its impact has even been broader as it has advanced the use of stochastic processes and the study of dynamical fluctuations in physical phenomena and far beyond. The central idea is atomism, that the surrounding heat bath consists of many fast moving atoms or molecules colliding with the particle, simultaneously  adding friction and noise.  Under the systematics of the van Hove or weak coupling limit, or as an application of the Zwanzig-Mori formalism, a Markov process is obtained for the particle motion suspended in that equilibrium environment \cite{chan,zwa,vanh}.   The resulting diffusion satisfies the fluctuation--dissipation relation as a result of time-reversal invariance of the underlying Hamiltonian dynamics for system and environment together.  It implies that the induced motion satisfies detailed balance.\\
A similar strategy can be applied when particles are in contact with spatially well-separated equilibrium reservoirs, mechanical, chemical or thermal as is the case.  Then and again in the suitable scaling limit, \emph{local} detailed balance governs the fluctuating dynamics of the system \cite{hal,derrida,leb,time}.  It allows a physically reasonable ground to study transport properties and fluctuations in the created currents.  The revival of nonequilibrium statistical mechanics during the last two decades has mostly concerned such systems.\\
The present paper tackles a third stage of that problem, where we consider a probe (or tracer) coupled  \emph{directly} to a nonequilibrium medium.  In many interesting cases that medium is active, where particle locomotion is coupled with internal or external nonequilibrium degrees of freedom. Typical examples include self-propelled bacteria or small beads, Janus particles, which can be optically or mechanically activated to run and tumble \cite{ram,mar1,mar2}. It is often an excess in dynamical activity, a nonphoretic driving or more generally, a breaking of the Einstein relation which results in such nonequilibrium behavior.  Various simplified models have been proposed for such media, and this paper takes the active Ornstein-Uhlenbeck (AOU) process for the dynamics of their particles.  The main specific question is to integrate out the AOU-particles coupled to a passive probe and thus to derive the induced motion of the probe. Interestingly, we will discover mass-generation on the probe from the persistence in the medium, and additional (possibly negative) friction breaking the Einstein relation. \\

The interest of benchmarking fluctuating motion in active environments  is  clear from the wealth of studies in experimental and theoretical microrheology and biophysics, witnessed e.g. by \cite{lib,pao,rold,gal,che,magi,ang,kou}. For example, the influence of active or critically driven droplets on e.g. transcriptional regulation has recently moved to the foreground \cite{hche,cho}. The general question is how nonequilibrium features in the active particles' motion are transferred to the probe dynamics.  Alternatively, probe motion can provide information about the condition of the active medium, and its analysis may lead to diagnosing  the quality or efficiency of life proceses.  But we first need to understand accurately how activity parameters such as the persistence play a role in the resulting fluctuating dynamics.  Such studies are also supported by the pleasant circumstance that possibilities of observation and manipulation of mesoscopic kinetics have been growing sensationally, thanks to e.g. optical tweezing and fast-camera tracking. Developments using a trajectory-based  approach to response  are in line  with these new tools.  Theoretical progress on the nature of the induced motion  has been reported in \cite{kafri,pao,kafri2,ray,ran,lei}, using a variety of approximations and/or numerical work to characterize interactions and induced forces.  To the best of our knowledge however, the fluctuating dynamics of one or more probes in an active bath has not been \emph{derived}.  We want  a constructive approach enabling to see in exactly what sense the probe gets activated. The methodology for such studies has been developed first in \cite{jsp,stefan} for deriving the motion in nonequilibrium media satisfying local detailed balance. Yet, in the present case, the translational motion of the active particles making the medium is nonMarkovian and far from even local detailed balance. 

\section{Active Ornstein--Uhlenbeck medium}
A commonly used model for an active overdamped dynamics is through the introduction of an Ornstein-Uhlenbeck noise; see \cite{mag,10,ot,fa,sza,magi,fod,wit,wit2,man}.  Its simplicity has allowed a rich variety of  simulation and theoretical studies, including confinement and interactions.  It has been studied in the context of  motility-induced phase transitions \cite{fa,mip} but also for modifications in glass physics \cite{oth2}, or other phenomena such as pressure on walls \cite{oth3} etc.  The active Ornstein-Uhlenbeck (AOU)-dynamics is written in \eqref{aou}--\eqref{oun} for coupling with the probe position. The activity resides in the colored noise $v(s)$ relecting the pushes from many hidden active components such as molecular motors.  In that sense, an  AOU-process is thought to  model the dynamics of tracers in living systems \cite{ben,fod2,fod3}. Here we use a probe to trace a medium of AOU-particles.\\

We refer to a {\it probe} now instead of to a particle while the constituents of the active medium are named {\it particles}.  After all, those active particles usually do not at all represent atoms, ions or molecules but are themselves (at least) micron-sized and in contact with an equilibrium environment of thermal or chemical baths for dissipating their waste energy.  The probe is even bigger than the active particles, could be a wall, with fluctuations visible in the dynamics for probes to tens of a micrometer in size \cite{lib}.  Nevertheless, our study remains conceptually a direct extension of the derivation of Brownian motion for a colloid suspended in a thermal equilibrium bath. The same assumptions are made concerning time-scale separation and weak coupling except for the equilibrium condition of the bath. Even local detailed balance is indeed violated in the active medium.  We will not use knowledge of the stationary distribution of the medium, even though many aspects are known for the simplest versions \cite{to,ot,oth4}.  Instead, we use a method, trajectory-based response theory \cite{resp}, which is both applicable to more complicated situations and is useful for obtaining the resulting fluctuating dynamics in explicit measureable quantities.

\section{Set up around quasistatic limit}
Denoting with $Y_t$ the position of the probe with mass $M$ at time $t$, its Newton equation (in one-dimensional notation always for simplicity) reads
\begin{equation}
M\ddot{Y_t} = -\lambda\sum_{i=1}^N(Y_t-x_i(t))\label{prm} 
\end{equation}
and couples with $N$ AOU-particles having positions $x_i(t)$.
The linearity of the coupling is not to be taken physically literal, but is only a mathematical simplification.  For an interaction potential $\Phi(|Y_t-x_i(t)|)$ with  short-range $\delta$, we should think of $\lambda = \rho\,\delta\,\Phi''(0)$  (positive or negative) with $\rho$ the medium-density, as measuring the effective spring constant.\\
  The AOU overdamped dynamics is  (for each particle) 
\begin{eqnarray}\label{aou}
\dot x(s) &=& {\cal E}\,v(s) + \lambda\mu\,(Y_s-x(s)) -\mu V'(x(s))\\
\tau\dot{v}(s) &=& -v(s) + \sqrt{2R} \, \xi(s)\label{oun}
\end{eqnarray}
Their activity depends on a persistence time $\tau$ and noise amplitudes ${\cal E}$ and $R$.  The $\xi(s)$ is standard white noise representing further hidden degrees of freedom.  For $\tau=0$, $v(s)=\sqrt{2R}\,\xi(s)$ can be substituted in \eqref{aou} and the medium becomes passive, characterized by an overdamped Langevin dynamics in contact with an equilibrium bath at temperature $T = {\cal E}^2 R/(k_B\mu)$ for mobility $\mu$. We use the same notation in the active case as well, even though $T$ no longer represents the physical temperature then. The potential $V=V_N$ on each AOU-particle may effectively represent a mean-field interaction and may depend on the density profile as well, which gives interesting possibilities for (e.g. motility induced) phase transitions \cite{mip}.  To make sure, other forces and noise can be added to the Newton equation \eqref{prm}, even to the extent of making the motion overdamped.  All those would be additive to \eqref{prm}.  We focus therefore on the main subject which instead is to integrate out the active medium from \eqref{prm}.\\
We proceed under the usual assumptions where the probe's motion is derived as an expansion around the quasistatic limit, also known as the Born-Oppenheimer approximation or adiabatic limit; see e.g. \cite{vanh,chan,zwa}.  It means that the probe is much slower with respect to the AOU-particles enabling a study in the spirit of Einstein--Laub theory for probes in electromagnetic media.  Secondly, we assume a weak coupling $\lambda$, possibly scaling with $N$ and the duration of the coupling to ensure compatibility with the quasistatic regime.\\
As quasistatic reference process we take the dynamics 
\begin{equation}\label{qaou}
\dot x(s) ={\cal E}\,v(s) + \lambda\mu\,(Y_t-x(s)) -\mu V'(x(s)),\quad s\leq t\\
\end{equation}
which is \eqref{aou} but for held-fixed probe position $Y_t$.  The OU-noise $v(s)$ still works from \eqref{oun}. With \eqref{qaou}, the resulting $x(s)-$process enjoys a nonequilibrium nonMarkovian steady condition for which we denote the statistical averaging by $\langle \cdot \rangle_\tau^\lambda$, of course also depending on the instantaneously-static probe position $Y_t$ except for $\lambda=0$. 
 Informally, the $s-$time of the AOU-particles in \eqref{qaou} runs much faster than the $t-$time of the probe. In that quasistatic approximation, \eqref{prm} becomes 
\begin{equation}
M\ddot{Y_t} = -\lambda N\,\left(Y_t-\langle x\rangle_\tau^\lambda\right)\label{prmq} 
\end{equation}
as if the active particle positions have relaxed to their stationary expectation for the instantaneous probe position $Y_t$.
We use an expansion around the reference dynamics \eqref{qaou}--\eqref{prmq} to integrate out the AOU-medium from \eqref{prm}.  

\section{Result}
The main result  can be summarized in the following structure for the induced probe's motion.  Suppose the original mass of the probe is rescaled as $M =\lambda_0 N\, m$ with $m$ a reference mass and $\lambda = g\lambda_0$ with $\lambda_0$ dimensionless.  Integrating out the AOU-medium from \eqref{prm}, we find
\begin{equation}\label{1sm}
(m + g\,\tau^2\frac{\lambda\,\mu}{2a})\,\ddot{Y_t} = -g\left(Y_t -\langle x\rangle_\tau^\lambda\right)
-\gamma\, \dot{Y}_{t} + \sqrt{2\Gamma}\,\xi_t
\end{equation}
with rescaled mass shifted 
proportional to $\tau^2$ (persistence) and  the coupling $a^{-1} = \int_0^\infty \id s\langle x(s)\,;\,v(0)\rangle^0_\tau/({\cal E}R)$ between spatial displacement and OU-noise in the medium.  The friction in \eqref{1sm} 
\begin{equation}\label{2fr}
\gamma=\frac{\Gamma}{2T} + \frac{g\lambda\mu}{2T}\,\int_0^{\infty}\id s\,s\,\langle x(s)\,;\,V'(x(0))\rangle_\tau^0
\end{equation}
connects its first term with the white noise amplitude $\Gamma = g\lambda\,\int_0^{\infty}\id s\,\langle x(s)\,;\,x(0)\rangle_\tau^0$.  
The second term  carries no definite sign and depends on the correlation between the effective force on an AOU-particle and its displacement.  The mass shift and the extra contribution to the friction are the main changes due to the activity.\\  In \eqref{1sm} we took a Markov approximation of the following more accurate description of the induced probe motion:
\begin{eqnarray}
&&M\ddot{Y_t} = -\lambda\,N\,Y_t + \lambda N \langle x\rangle_\tau^\lambda\label{1s}\\
&& - \frac 1{2}\lambda^2N\,\int_0^{\infty}\id s\,K(s)\, \dot{Y}_{t-s} + \sqrt{\lambda^2N}\,\eta_t\label{fors}\\
&&- \frac 1{2}\lambda^2N\int_0^{\infty}\id s\,  \,{\cal D}(s)\,\frac{Y_t-Y_{t-s}}{s}\;\;{   }\label{trs}\\
&&- \tau^2\,{\cal E}\,\frac{\lambda^2N}{2T}\int^{\infty}_0\id s\;\ddot{Y}_{t-s}\,\left<x(s)\,;\,v(0)\right>_\tau^0 \label{pas}
\end{eqnarray}
We ignore terms of order $O(\lambda^3N)$. All expectations $\langle\cdot\rangle_\tau^\lambda$ refer to averages over the stationary AOU-medium defined from \eqref{qaou}.   The first line \eqref{1s} reflects the quasistatic limit \eqref{prmq}, of leading order
 for small $\lambda$.   We come next to the physical interpretation of the additional terms.\\
In \eqref{fors} the friction kernel 
\begin{equation}\label{frk}
K(s) = \frac{1}{T}\,\langle x(s)\,;\,x(0)\rangle_\tau^0 = \frac 1{T}\,\langle \eta_s\,;\,\eta_0\rangle_\tau^0
\end{equation} 
is proportional to the noise covariance as in the standard fluctuation--dissipation relation (FDR).  The noise $\eta_t$ is stationary, has mean zero and becomes Gaussian white noise under the same conditions as in the passive equilibrium case, \cite{chan,zwa,vanh}.  It picks up the distribution of 
\begin{equation}\label{nonos}
\eta_s = \frac 1{\sqrt{N}}\sum_{i=1}^N \left(x_i(s) -
\langle x\rangle_\tau^\lambda\right)
\end{equation}
in the AOU-medium following \eqref{prm}.  So far, lines \eqref{1s}--\eqref{fors} would also be present for a passive medium $\tau=0$; we say more about the factor 1/2 below.\\
 The third line \eqref{trs} is an additional friction, with kernel
 \begin{equation}\label{ndf}
 {\cal D}(s) = \frac{\mu}{T}\,s\,\langle x(s)\,;\,V'(x(0))\rangle_\tau^0
\end{equation}
which is unbalanced by the noise. Note that $(Y_t-Y_{t-s})/s\simeq \dot{Y}_{t-s}$ for small $s$, which is most relevant as ${\cal D}(s)$ is expected to decay exponentially fast; see \eqref{linc}.
As we show in Eq. S9 of the Appendix, in the passive case $\tau=0$,
\begin{equation}
\int_0^{\infty}\id s\,{\cal D}(s)\,\frac{Y_t-Y_{t-s}}{s}
=  \int_0^{\infty}\id s\,K(s)\, \dot{Y}_{t-s} \label{pase}
\end{equation}
which then removes the factor 1/2 from \eqref{fors}, to install the standard FDR from \eqref{frk}.
In the active case however, \eqref{trs} is the explicit source of violation of the FDR in the induced fluctuating dynamics.   We call \eqref{ndf} the non-dissipative friction kernel as \eqref{trs} corresponds to the so called frenetic contribution in linear response, of the same order as the dissipative friction with kernel \eqref{frk}; see \cite{fren,resp}.  The kernel \eqref{ndf} may produce a negative contribution to the friction. Indeed, its sign is decided from the time-correlation between AOU-position $x(s)$ and the effective force $-V'(x(0))$.   Imagine for example the  (many) $N$ particles, each obeying \eqref{aou} but replacing there $V\rightarrow \Phi$ describing interactions, 
\[
\Phi(x_1,\ldots,x_N) = \sum_{i,j=1}^N W_N(x_i,x_j) + \sum_{i=1}^N\phi(x_i)
\]
For such interacting dynamics, the AOU-particles may show a
motility-induced phase separation \cite{10,mip}. A mean field version is readily obtained when $W_N \propto W/N$ with $W(x,y) = W(x) + m_1 W'(x) + m_2 W''(x)$ independent of $y$ and with $m_i$ describing effectively the density and its spreading.  A typical choice in \eqref{aou} would then be to take
\begin{equation}\label{mff}
V'(x) = \phi'(x) + W'(x) + m_1 W''(x) + m_2 W'''(x)
\end{equation}
making an active Ornstein-Uhlenbeck process \eqref{aou} where the interaction and density are parameterized via the parameters $m_1,m_2$.  The different terms in \eqref{mff} may naturally produce different signs in \eqref{ndf}. In particular it is allowed to have a contribution $k\,\langle x(s)\,;\,x(0)\rangle^0_\tau < 0$ for $k<0$ in \eqref{ndf} without breaking the stability of the AOU-process. It is well known that for active particles trapping may lead to clustering at the boundaries of the trap, which is an effective repulsion from the center of the trap \cite{tail}. Heuristically, it should also not be too surprising to imagine that persistent flocking carries the probe, transporting energy from the active medium to the probe.   
 Negative differential mobility has been found before due to frenetic causes in  e.g. \cite{frec,sar}. \\
The final line \eqref{pas} is explicit in the persistence $\tau$ and introduces extra inertia with a memory-kernel for extra mass to the probe. The bare mass $M$ of the probe gets modified via the time-correlation of the position $x(t)$ and the free OU-noise $\cal E\,v(s) =\dot x(s) + \mu \,V'(x(s))$, as summarized below in \eqref{ms}.  
 That can be seen as an instance of entropy production (due to the active medium) renormalizing the inertial mass of the immersed probe.  A similar effect has been observed in the stabilization of a probe when in contact with a driven environment \cite{cyc}.\\  
 
 In order to understand the dependence of the time-correlations appearing in \eqref{1s}--\eqref{trs} on the persistence $\tau$, we have calculated them for the fully linear case where $V(x) = kx^2/2$ for  spring constant $k$.  Details are presented in Section 2 of the Appendix. With $a=\mu(k +\lambda)$, the explicit results  are
 \begin{eqnarray}
 \langle x(s)\,;\,x(0)\rangle^\lambda_\tau &=& \left[ \tau\,e^{-s/\tau} - \frac 1{a}e^{-as} \right]  \frac{R{\cal E}^2}{a^2\tau^2-1}\label{linc}\\
 \langle x_s\,;\,v_0\rangle^\lambda_\tau &=& \frac{R\,{\cal E}}{a\tau-1}\,\left\{e^{-s/\tau}-e^{-as}\frac{2}{a\tau+1}\right\}\label{lincv}
 \end{eqnarray}
Correlation \eqref{lincv} gives the behavior for the kernel in \eqref{pas}.  For large $a\tau \gg 1$ (far-from-equilibrium regime), we have $ \langle x_s\,;\,v_0\rangle^0 \rightarrow \frac{R\,\cal E}{a}\,\frac1{\tau}e^{-s/\tau}$ which, according to \eqref{pas}, means that the inertial mass of the probe shifts as
\begin{equation}\label{ms}
M\rightarrow M + \tau^2\,{\cal E}\,\frac{\lambda^2N}{2T}  \frac{R\cal E}{a} = M + \tau^2\frac{\lambda^2N}{2k} 
 \end{equation}
 from the influence of the (highly-persistent) active medium. The explicit result \eqref{ms} is for the correlation \eqref{lincv} in the linear AOU-system but the interpretation is wider: the mass shift is proportional to the persistence over the trapping, which is typically a measure of the variance in the position of active particles. \\
For intermediate regimes, there are two time-scales in the kernels, determined by $\tau$ and $a^{-1}$. As already mentioned,
for small persistence we recover the standard Langevin equation in the limit $\tau \ll a^{-1}$.

\section{Derivation of the fluctuation dynamics}
The slowly moving probe provides a stimulus to the AOU-medium. The first correction around the quasistatic limit  \eqref{qaou}--\eqref{prmq} is provided by the linear response of the medium; see \cite{resp} for an introduction.  The response or back-reaction of the AOU-particles determines the fluctuating motion of the probe \cite{jsp,stefan,leipzig,krueger}.   The linear response around AOU is treated as Section 1 in the Appendix.\\
We compare two AOU-ensembles, one where the path $(Y_s, s\leq t)$ of the probe is given, and one where the probe is held fixed at the final $Y_t$. As we treat the AOU-particles effectively independent, it suffices to investigate the coupled dynamics of just one AOU-particle with the probe. We compare the two ensembles by introducing an action $\cal A (\omega)$, as a function of the AOU-particle trajectories $\omega = (x_s, s\leq t)$, and (not indicated) depending on probe path $(Y_s)^t= (Y_s, s\leq t)$: 
\begin{equation}\label{app}
\text{Prob}(\omega|Y_s, s\leq t) = \exp [-{\cal A}(\omega)]  \;  \,\text{Prob}(\omega|Y_s=Y_t, \text{ for all } s\leq t) 
\end{equation}
The left-hand side gives the weight on trajectories where the AOU-particle follows \eqref{aou}, i.e., is conditioned  on $(Y_s)^t$. The probability in the right hand-side is the reference weight for AOU-particle trajectories supposing the probe has always been at the (final) position $Y_t$.  The action $\cal A$ in \eqref{app} is entirely explicit as computed in Section 1 of \cite{sm},
\begin{equation}\label{a}
{\cal A}(\omega) = -\frac 1{2\cal E^2}\int\id s\, {\cal K}(s)\,\big( \dot x(s) + \mu U'(x(s))\big)
\end{equation}
where the potential is $U(x) = V(x) + \lambda(Y_t-x)^2/2$ and the kernel equals 
\begin{equation}\label{kern}
{\cal K}(s) =  \frac{\lambda\mu}{R}\,(Y_s-Y_t -\tau^2 \ddot{Y}_s)
\end{equation}
function of the difference in probe position $Y_s-Y_t$ at time $s$.  The persistence time $\tau$ enters crucially in the kernel (and action) as $\tau^2$ multiplying $\ddot Y_s$.\\
The relation \eqref{app} implies $\langle x(t)\,|\,(Y_s)^t\rangle = \langle x(t)\rangle^\lambda_\tau - \langle x(t)\,;\,{\cal A}\rangle^\lambda_\tau$ to first order in $Y_s-Y_t$.  With \eqref{a} we thus obtain
the average force of each AOU-particle (leaving away subscripts) on the probe, given the probe path $(Y_s,s\leq t)$,
\begin{equation}\label{alinr}
\langle x(t)\,|\,(Y_s)^t\rangle =  \langle x(t)\rangle^\lambda_\tau +  \frac{1}{2\cal E^2}\int^t_{-\infty}\id s\, {\cal K}_s\,\left<x(t)\,;\,\big( \dot x(s) + \mu U'(x(s))\big) \right>^\lambda_\tau 
\end{equation}
The covariances $\langle\cdot\,;\,\cdot\rangle_\tau^\lambda$ are as before in the quasistatic process \eqref{qaou}. Taking small $\lambda\mu\,(Y_s-Y_t)$ means that the probe has not wandered away too far in the past $s\leq t$ over a time $(\lambda\mu)^{-1}$ which combines the coupling to the probe with the mobility of the AOU-particles.  All terms in \eqref{fors}--\eqref{trs}--\eqref{pas} except the noise follow from \eqref{alinr} by substituting \eqref{kern}: for example,
\begin{eqnarray}\label{fterms}
&&\frac{\mu}{R\cal E^2}\,\int^t_{-\infty}\id s\, (Y_s-Y_t)\,\left<x(t)\,;\,\big( \dot x(s) + \mu V'(x(s))\big) \right>^\lambda_\tau=\nonumber\\
&&-\int_0^{\infty}\id s\,\left[ K(s)\, \dot{Y}_{t-s} + {\cal D}(s)\,\frac{Y_t-Y_{t-s}}{s}\right]
\end{eqnarray}
give the friction terms.
The active contribution in the kernel \eqref{kern}  gives
\begin{equation}\label{pent}
\frac{\lambda\mu\tau^2}{2R{\cal E}^2}\,\int^t_{-\infty}\id s\, \ddot{Y_s}\,\left<x(t)\,;\,\dot x(s) + \mu \,U'(x(s))\right>^\lambda_\tau 
\end{equation}
where $\dot x(s) + \mu \,U'(x(s)) = \cal E v_s$ in the quasistatic process \eqref{qaou}.  It gives \eqref{pas} in the induced probe dynamics.  
To summarize, \eqref{fterms} and \eqref{pent} give the force on the probe by each active AOU-particle in \eqref{prm} such as decomposed in \eqref{alinr} (up to a factor of $\lambda$).
The corresponding terms in \eqref{fors}--\eqref{trs}--\eqref{pas} follow from replacing $U\rightarrow V + O(\lambda)$.\\

Let us finally consider the noise.  The force in \eqref{prm} by each AOU-particle on the probe is fluctuating as
\begin{eqnarray}
\lambda\,x(t) &=&  \lambda \,\langle x(t)\,|\,(Y_s)^t\rangle + \zeta_t \label{fobi} \\
 \zeta_t &=&  \lambda\,x(t)-\lambda \,\langle x(t)\,|\,(Y_s)^t\rangle \label{nois} 
\end{eqnarray}
where $\zeta_t$ is a ``noise''  that depends on the AOU-medium and that has mean zero for every probe trajectory $(Y_s)^t$. The noise $\zeta_t$ enters (in a rescaled way) as $\eta_t$ in \eqref{fors}. Its mean $\langle \zeta_t\rangle =0$ and its covariance is
\begin{equation}\label{noi}
\langle\zeta_s\,;\,\zeta_{s'}\rangle = \lambda^2\,\langle x(s)\,;\,x(s')  \rangle^0_\tau 
\end{equation}
in leading order:
\[
\frac{\mu}{{\cal E}^2R}\,\langle\zeta_s\,;\,\zeta_{s'}\rangle = \lambda^2\,K(s-s')
\]
which is the remnant \eqref{frk} of the Einstein relation (taking $k_B=1$).
As true in the passive case as well, the noise need not be determined by its second moment.  Whether it becomes Gaussian or even white depends on the steady AOU-medium and on performing the full weak coupling limit. No additional complications enter as long as the AOU-medium allows sufficient decay of spacetime correlations for the central limit theorem to work.

\section{Conclusions and outlook}
Benchmarking the fluctuating dynamics of a probe immersed in an active medium is an important step beyond the Brownian motion for tagged particles suspended in equilibrium baths. An expansion around the quasistatic limit is possible by an application of linear response theory around a nonequilibrium steady condition. The frenetic contribution adds to the dissipative part in the response and produces extra terms in the induced probe dynamics.  A first term modifies the effective mass of the probe and is proportional to the persistence time.  Persistence is inherited by the probe from its interaction with the active medium, which effectively generates extra inertia for the probe. The second term breaks the Einstein relation  by giving a contribution to the friction which is not compensated by the noise.  In general that ``extra'' friction depends on  the mutual forces between the medium-particles and their correlation with the probe position as in the modified Sutherland--Einstein relation \cite{proc,sar} and in the Harada-Sasa equality \cite{har}.\\

The derivation in the present paper has been restricted to active Ornstein-Uhlenbeck particles making the medium.  That is the easiest for being explicit in the response formula.  The structure of the derivation strongly suggests that the same  properties and modifications will be found in the induced probe dynamics for more general active media.  We look forward to experimental tests where both the mass-renormalization and the extra (possibly negative) friction would be observed.  Of substantial interest is to discover the influence of the motility induced phase transitions in the media on the fluctuating probe motion.  In the present analysis, that was modeled effectively via the potential $V$ in \eqref{mff}.

\appendix

\section{Linear response for an AOU-particle}
To understand the method of deriving the induced dynamics Eqs 8--11 in the main text we turn to response theory for AOU-particles.  We rewrite the original model Eq 2 for $s\leq t$ as
\begin{equation}\label{aoup}
\dot x(s) = \cal E\,v(s)  - \mu U'(x(s)) + h_s
\end{equation}
where the force $h_s = \lambda\mu(Y_s-Y_t)$ is interpreted as a time-dependent perturbation (independent of $x(s)$) and the potential $U(x) = V(x) + \lambda(Y_t-x)^2/2$.   The perturbation $(h_s)$ measures the distance between $Y_s$ and $Y_t$, i.e., around the quasistatic limit Eq 4 where $h_s=0$.\\
Linear response for AOU-particles has been considered before in \cite{pao,cip,resAB} but we use a different approach.
To organize the response theory for such a nonequilibrium process, we use dynamical ensembles \cite{resp,fdr}. The perturbed ensemble Prob$^h[\omega] = e^{-\cal A}\,$Prob$[\omega]$ has a density governed by an action $\cal A = \cal A(\omega)$ on trajectories $\omega$ of positions $x(s), s \leq t,$ and on the protocol $(h_s)$.  That is also the starting point of the functional calculus method of Fox \cite{fox} but the subsequent application is totally different.\\   
While the OU-noise $v(s)$ (in Eq 3 of the main text) is not white, it is Gaussian. Therefore, the probability of a trajectory $\omega$ of positions $x(s)$, is proportional to
\begin{equation}\label{ker}
\text{Prob}^h[\omega] \propto \exp - \frac 1{2} \int \id s \int\id s'\,\Gamma(s-s') v(s) v(s')
\end{equation}
where we must substitute
\[
v(s) = \frac 1{\cal E}\left( \dot x(s) + \mu U'(x(s)) - h_s \right)
\]
and with symmetric kernel $\Gamma(s)$ for which
\begin{equation}\label{con}
\int \id s'\,\Gamma(s-s') \,\left< v(s')v(s'')\right> = \delta(s-s'')
\end{equation}
Since the noise-correlation has exponential memory,
\begin{equation}\label{ga}
\langle v(s)v(s')\rangle = \frac{R}{\tau}\,e^{-|s-s'|/\tau}  
\end{equation}
from \eqref{con} we find  $\Gamma(s) = [\delta(s) - \tau^2 \ddot\delta(s)]/(2R)$ as solution.  
As a result, since  the reference ensemble has $h_s\equiv 0$, to linear order the action is 
\begin{equation}\label{aca}
{\cal A}(\omega) = -\frac 1{2\cal E^2}\int\id s\, {\cal K}(s)\,\left( \dot x(s) + \mu U'(x(s))\right)
\end{equation}
where the kernel equals ${\cal K}(s) =  2
\int \id s'\, h_s'\, \Gamma(s-s') = h_{s}/R - \tau^2 \ddot{h}_{s}/R$. 
The AOU-model (for $\tau>0$) does not show short-time diffusion, and hence there is no difference here between e.g. Stratonovich and It\^o conventions for the stochastic integral.\\

Coming to linear response, we consider the dynamics \eqref{aoup} and we wish to evaluate how single-time observations (say, via a function $f$) get modified with respect to the nonequilibrium reference dynamics where $h_s\equiv 0$: in terms of the average $\langle\cdot\rangle^0$ over the unperturbed dynamical ensemble, with $\omega_t=x(t)$,
\begin{eqnarray}
\langle f(x(t))\rangle^h - \langle f(x(t))\rangle^0 &=& 
\int {\cal D}[\omega]\,\text{Prob}^0[\omega] \,f(\omega_t)[e^{{-\cal A}(\omega)} -1]\nonumber\\
&=& -\int {\cal D}[\omega]\,\text{Prob}^0[\omega] \,f(\omega_t){\cal A}(\omega) = -\langle f(x(t))\,;\,{\cal A}(\omega)\rangle^0\label{dyan}
\end{eqnarray}
where the action $\cal A$ is given in \eqref{aca}, normalized as $1 =\int {\cal D}[\omega]\,\text{Prob}^h[\omega] = \int {\cal D}[\omega]\,e^{-\cal A}\,\text{Prob}[\omega]$, and therefore satisfying
\begin{equation}\label{norm}
\langle e^{{-\cal A}}\rangle^0 = 1 \implies \langle \cal A\rangle^0 = 0
\end{equation}
to linear order in $h$. That is why we can take the covariance in \eqref{dyan}, denoted by $\langle A\,;B\rangle^0 =\langle AB\rangle^0 - \langle A\rangle^0\langle B\rangle^0$.\\
From \eqref{aca}, \eqref{dyan} continues as
\begin{equation}\label{plinr}
\langle f(x(t))\rangle^h - \langle f(x(t))\rangle^0 =\frac 1{2\cal E^2}\int_{-\infty}^t\id s\, {\cal K}_s\left< f(x(t))\,;\,\big( \dot x(s) + \mu U'(x(s))\big) \right>^0
\end{equation}
We need $f(x)=x$ and $\langle x_t \,;\,{\cal A}\rangle^0$ for the response in the main text.   All that explains Eqs 20--23 there.\\

Note that \eqref{norm} implies that 
\[
\int\id s\, {\cal K}(s)\,\big< x(0)\,;\, \dot x(s) + \mu U'(x(s)) \big>^0= 0\]
When $\tau=0$ there is time-reversal symmetry, and hence, in that passive case,
\begin{equation}\label{pasrev}
\int\id s\, {\cal K}(s)\,\big< x(s)\,;\, -\dot x(0) + \mu U'(x(0)) \big>^0= 0
\end{equation}
which is Eq 15 in the main text.

\section{Linear AOU-process}
For explicit calculations we change the dynamics from \eqref{aoup} to
\begin{eqnarray}\label{lao}
\dot{x}_s &=& {\cal E } v_s - a \, (x_s-b)  + h_s \nonumber \\
\dot{v}_s &=& -\gamma\,v_s + \sqrt{2D} \, \xi_s
\end{eqnarray}
with \[
\gamma=\frac 1{\tau}, \,D = R /\tau^2 = R\gamma^2,\;\; a=\mu(k+\lambda),\; b=\frac{\lambda\mu}{a}Y_t,\;h_s= \lambda\mu\,(Y_s-Y_t)
\]
making the translation to the notation in the main text. By the linearity, the dynamics can be solved completely, giving
\begin{equation}
x_t = b+ (x_0-b)\,e^{-at} + \int_{0}^t \id s\,e^{-a(t-s)} [h_{s}  + {\cal E}\, v_s]\label{nao}.
\end{equation}
for
\[v_s=v_0e^{-\gamma s} +\sqrt{2D}\int_{0}^s e^{-\gamma(s-u)}\,\xi_{u} \,\id u\]
or also,
\begin{eqnarray}
x_t &=& b+  (x_0-b)e^{-at} \label{lins}\\
+&& \int_0^t
\id s e^{-a(t-s)} [h_s + {\cal E} v_0 e^{-\gamma s}] + \frac{\sqrt{2R{\cal E}^2}}{1-a\tau} \int_0^t\id s\, \xi_s[e^{-a(t-s)}-e^{-\gamma(t-s)}]\nonumber
\end{eqnarray}
From here we can compute the steady state correlation functons, as also obtained in Section IVB of \cite{sza} (but correcting here a typo in $\langle (x_0-b);v_t\rangle^0$),
\begin{eqnarray}\label{cors}
&& \langle (x_t-b);(x_0-b)\rangle^0 = \left[ \tau\,e^{-t/\tau} - \frac 1{a}e^{-at} \right]  \frac{R{\cal E}^2}{a^2\tau^2-1}\\
&& \langle (x_0-b);v_t\rangle^0 = e^{- t/\tau}\, \frac{R{\cal E}}{a\tau+1}\\
&& \langle (x_t-b);v_0\rangle^0 = \frac{R\,{\cal E}}{a\tau-1}\,\left\{e^{-t/\tau}-e^{-at}\frac{2}{a\tau+1}\right\}\\
&& \langle (x-b)^2\rangle^0 = \frac{R\,{\cal E}^2}{(a\tau+1) a}\\
&& \langle v_0\,;\,v_t\rangle = \frac{R}{\tau}\,e^{-t/\tau}
\end{eqnarray}
Those are the correlations used in Eqs 17-18 of the main text.  Note that having $k<0$ is no problem as long as $a>0$.  That refers to the comments below Eq 16.\\
As a final observation we emphasize that while \eqref{nao}--\eqref{lins} are exact, the decomposition  in Eq 6 or in Eqs 8--11 is physically more useful.  In  particular,  the noise in \eqref{lins} does not have the simple representation of Eq 13 in the main text.  Obviously, going nonlinear  remains compatible with the perturbative approach around the quasistatic limit anyhow.

\end{document}